\documentclass[10pt,aps,pre,superscriptaddress,twocolumn,showpacs,floatfix,noeprint]{revtex4-1}
\usepackage[utf8]{inputenc} % Para usar com codificacao de caracter utf8
\usepackage{mathtools}
\usepackage{amsfonts}
\usepackage{amsmath}
\usepackage{amsthm}
\usepackage{subfigure}
\usepackage{graphicx}
\usepackage{epsfig}
\usepackage{multirow}
\usepackage{rotating}
\usepackage{tikz}
\usepackage{pgfplots}
\pgfplotsset{compat=1.16}
\newcommand{\IFUFG}{Instituto de F{\'i}sica, Universidade Federal de 
Goi{\'a}s, Av. Esperan\c{c}a s/n, 74.690-900, Goi{\^a}nia, GO, Brazil}

\newcommand{\IFG}{Instituto Federal de Goi{\'a}s, Rua 76, Centro, Goi{\^a}nia - GO, Brazil}
\begin{document}

\title{2D triangular Ising model with bond phonons: An entropic simulation study}

\author{R. M. L. Nascimento}
\affiliation{\IFUFG}
\author{Claudio J. DaSilva}
\affiliation{\IFG}
\author{L. S. Ferreira}
\affiliation{\IFUFG}
\author{A. A. Caparica}
\affiliation{\IFUFG}
\email{caparica@ufg.br}

\begin{abstract}
In this work, we study and evaluate the impact of a periodic spin-lattice coupling in an Ising-like system on a 2D triangular lattice. Our proposed simple Hamiltonian considers this additional interaction as an effect of preferential phonon propagation direction augmented by the symmetry of the underline lattice. The simplified analytical description of this new model brought us consistent information about its ground state and thermal behavior, and allowed us to highlight a singularity where the model behaves as several decoupled one-dimensional Ising systems. A thorough analysis was obtained via entropic simulations based in the Wang-Landau method that estimates the density of states $g(E)$ to explore the phase diagram and other thermodynamic properties of interest. Also, we used the finite size scaling technique to characterize the critical exponents and the nature of the phase transitions that, despite the strong influence of the spin-lattice coupling, turned out to be within the same universality class as the original 2D Ising model.
\end{abstract}

\maketitle

\section{Introduction}

One of the most important techniques for studying nature is abstraction, since the representation of all the characteristics of an observable is unfeasible and often does not add advances in the understanding of the properties of interest. Thus, modeling appears as a minimal representation of the characteristics of interest. An example is the Ising model \cite{Ising1925} which characterizes a magnetic material through a static lattice with magnetic moments at the vertices of the lattice. Furthermore, it only considers the $z$ direction of the magnetic moment. Despite the immense simplification, the model is capable of mimicking the main characteristics of a magnetic material, serving as a base model to understand various properties of magnetic materials. Due to its simplicity, practicality and efficiency, several other areas of knowledge base their models on this simplified version.

As the questions become more elaborate, the abstraction must take on characteristics closer to reality. In particular for magnetic models, there are several modifications of the Ising model to meet the needs of the materials, such as the Heisenberg, XY, Potts \cite{Potts1952}, Baxter-Wu \cite{Baxter1973}, $J1-J2$ models and many others. Each model has its own characteristics and some of them are still under intense study in order to be fully understood \cite{Jorge2021,Das2022}.

The main difference between these models lies in the interaction between neighbors (exchange interaction). The vibrations of the atoms are not considered, since the spins are fixed at the vertices of the lattice. A more realistic approach is to consider that atoms can oscillate around an equilibrium position and that normal modes of vibration can dynamically change the interaction between their neighbors. In this context, Oitmaa (1975) \cite{Oitmaa1975} investigates the effect on the exchange interaction caused by an elastic wave on the crystalline lattice. His results point to changes in the critical temperature and changes in the type of phase transition.

Since then, several other representations have been proposed with the aim of making the models increasingly realistic \cite{Bursill1999,Yin2013,Li2021}. However, the price paid for this is proportional to the level of reality required \cite{Ma2008,Mcguire2017,Choe2021,Mankovsky2022}. In particular, we can mention works that use the Heisenberg model taking into account the interaction of the spin with the lattice, where atoms can move around an equilibrium position, affecting the exchange interaction. Furthermore, it is considered an additional energetic term to the Hamiltonian due to its vibration \cite{Jia2005,Wang2008,Aoyama2021}. This additional energy, called elastic energy, classifies the models into bond-phonon and site-phonon \cite{Penc2004,Bergman2006,Wang2008,Aoyama2016}, which consider the effect of joint vibration and the effect of local vibration, respectively.

Aoyama \textit{et. al.} \cite{Aoyama2021} investigated theoretically effects of the magnetic field on the spin-lattice-coupled (SLC) ordering in the breathing-pyrochlore antiferromagnets, based on the two possible simplified models describing the SLC, the bond-phonon and site-phonon model. Their results show that the site-phonon model presents long-range ordering while in the bond-phonon model the vector-multipole orders, spin-liquid plateau and nematic phases are observed. Although the phases are different for the two representations of elastic energy, the phase diagrams are qualitatively equivalent.

The study of these models is conducted by Monte Carlo simulations using the Metropolis algorithm, because the translational and spin degrees of freedom are considered, a fact that demands high computational cost. Theoretical and entropic sampling work would pay an even higher price. Given this, we intend to propose a spin lattice model that considers the spin-lattice interaction and that is simple enough so that theoretical and simulation work can be carried out more easily.

In this direction, Pili and Grigera \cite{Pili2019} propose a model based on the Ising model with Einstein site phonons. They found that the coupling of the magnetic to the elastic degrees of freedom gradually lowers the magnetic ordering transition until it is completely suppressed at a critical value of the coupling constant. Above this the system suffers a simultaneous magnetic and structural transition into a dimerized state with lower crystalline symmetry and ferromagnetic clusters antiferromagnetically aligned. However, the use of the Metropolis algorithm makes a more detailed study of phase transitions unfeasible, leaving questions to be answered, such as the order of the phase transition and the critical exponents in the different regions.

To overcome these limitations, we intend to use entropic sampling to study the critical phenomena of a lattice model with spin-lattice interaction in a triangular lattice, estimating the critical exponents and the critical temperature. This technique, based on the Wang-Landau algorithm \cite{Wang2001}, has proven to be a powerful method for obtaining the thermodynamic properties of the system and determining the critical temperature and critical exponents \cite{Fytas2011,Gai2013,Caparica2015b,Jorge2019}. Furthermore, the estimation of the joint density of states \cite{Zhou2006} allows the study of critical phenomena for any value of the energetic parameters, easily obtaining the phase diagram of the system \cite{Silva2006,Jorge2021,Ferreira2021}.

We start this paper by describing the proposed model in section \ref{model}. In section \ref{comp}, we address the computational details used to obtain the results. Then, in section \ref{results}, we present the thermodynamic properties and discuss the main characteristics of the influence of lattice vibration on the triangular Ising model. Finally, in section \ref{conclusion} we summarize the results and allude to some final remarks.

\section{The model}\label{model}

In this section we will show the details of the model construction, from a more complete model to a simplified version. In general, to deal with the coupling of the spin with the lattice, the Heisenberg model is used plus the possibility of atoms being able to oscillate around their equilibrium positions and elastic energy \cite{Jia2005,Wang2008,Aoyama2021}. This vibration alters the exchange interaction, since it is dependent on the distance between the spins, so that it can be written as a function of the equilibrium distance between the atoms plus their infinitesimal displacements, that is, $J_{ij} (\vert \mathbf{r}_{ij }^{0} +\mathbf{u}_i - \mathbf{u}_j \vert)$, where $\mathbf{r}_{ij}^{0} $ is the equilibrium distance between two atoms and $\mathbf{u}_i$ ($\mathbf{u}_j$) is the displacement vector of atom $i$($j$) from the equilibrium position. Furthermore, we must guarantee that $\vert \mathbf{u}_i \vert / \vert \mathbf{r}_{ij} ^{0} \vert \ll 1$, known as the condition of small oscillations, which allows us expand the exchange interaction around the equilibrium position, obtaining

\begin{equation}\label{math03} 
 J_{ij} = J_{ij}( \vert \mathbf{r}_{ij}^{0} \vert ) + {\frac{dJ_{ij}}{dr}\lvert}_{r=\vert \mathbf{r}_{ij}^{0} \vert } \mathbf{e}_{ij} \cdot \delta \mathbf{u}_{ij}, 
\end{equation}
where $J_{ij}=J_{ij}( \vert \mathbf{r}_{ij}^{0} +\mathbf{u}_i - \mathbf{u}_j \vert )$, $\mathbf{ e}_{ij} \equiv \mathbf{r}_{ij}^{0} / \vert \mathbf{r}_{ij}^{0} \vert $ is the vector that connects two neighboring  $i$ and $j$ in their respective equilibrium positions and $\delta \mathbf{u}_{ij}=\mathbf{u}_i - \mathbf{u}_j$. The first term of this equation is a constant that depends only on the equilibrium distances, therefore it represents an undisturbed exchange interaction.
The second term depends on the directions of the displacements and the undisturbed direction of bonding between the atoms. For simplicity we define
\begin{equation}
     J_a\equiv \left(\frac{dJ_{ij}}{dr}\right)_{r_{ij}^0} |\delta \mathbf{u}_{ij}|
\end{equation}
as the intensity of the spin coupling with the lattice. It is this term that effectively changes the exchange interaction, since the undisturbed direction of bonds is constant.
\begin{figure}[!htb] 
\centering
\includegraphics[scale=0.75]{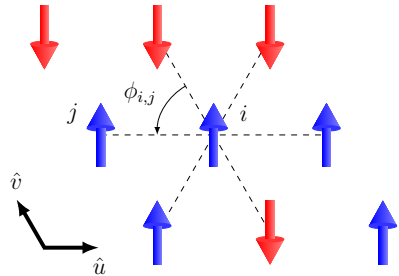}
\caption{Representation of a configuration in the triangular lattice. The dotted lines indicate the interactions between the first neighbors of the site $i$. Taking the direction given by the vector $\hat{v}$ as the main axis, the angle $\phi_{i,j}=$ 60\textdegree \, is represented in the figure.\label{fig01}}
\end{figure}

From this point of view, the vibrations of the atoms would be responsible for the coupling between the spin and the lattice, since they change the value of the exchange interaction. Alternatively, we can reverse the order and think that an intensity of the exchange interaction is associated with a certain vibration of the atoms. This change in perspective is fundamental on the way to simplifying the model. In this sense, the first simplification considered is that the atoms will remain in their equilibrium positions, that is, at the vertices of the lattice. The coupling of the spin with the lattice will be mediated by the value of $J_a$, so that the visualization of the vibrations of the atoms is abandoned. This assumption is reasonable, since the only dependence on the position of the atoms is in the vector $\delta \mathbf{u}_{ij}$, which, by definition, is much smaller than the equilibrium distance ${r}_{ ij}^{0}$. No lattice distortion information can be obtained from this simplification.

Furthermore, we consider that $\delta \mathbf{u}_{ij} ={u}_{ij}~\hat{v}$, where $\hat{v}$ is the versor that points in the direction of the main axis, thus it can be taken in any direction. We consider that it is in the direction of one of the lattice connections, as shown in Fig.\ref{fig01}. The choice of a main axis links the infinitesimal displacement of an atom to the infinitesimal displacement of its nearest neighbor, thus, we are considering a collective vibration. Thus we can say that the choice of the main axis is linked to the bond phonon model, even if an elastic energy is not explicitly added to the Hamiltonian. Linking $\delta \mathbf{u}_{ij}$ to the direction of the main axis does not allow the description of the distortions suffered by the lattice, since there is an infinite number of displacement vectors whose difference is in the direction of the main axis.

Finally, we restrict the spin degrees of freedom to the Ising model, so that the Hamiltonine for this model can be written as
\begin{equation}\label{math0}
\mathcal{H} =-\sum_{\langle i, j \rangle} J_{ij}\sigma_i\sigma_j,
\end{equation}
where $\sigma_i$ is the spin variable that can assume $\pm 1$ and the spin coupling-dependent exchange interaction with the lattice is
\begin{equation}\label{math04}
J_{ij} = J + J_a\cos(n\phi_{ij}),
\end{equation}
where $\phi_{ij}$ is the angle formed from the main axis and the line that joins the neighboring spins $i$ and $j$, as shown in the Fig.\ref{fig01}. In this sense, when we adopt a triangular lattice like in the Fig.\ref{fig01}, we observe that the angles formed between the main axis and the interactions between neighboring spins are multiples of $60^\circ$. The sum is performed over the first neighbors. We are interested in investigating the effects caused by the coupling of the spin with the lattice, so we will consider $J=1$.

Another important aspect of our proposal is the role of the variable $n$ that is responsible for ensuring the isotropy of bonds between two neighboring spins. We assume that $n$ is even, so that we can separate the possible values for $n\ge 2$ into two cases:
\begin{itemize}
    \item[I -] $n$ is a multiple of $6$
        \begin{itemize}
        \item [ ] we have $J_{ij}=J+J_a$ for all connections, so that, for $J+J_a>0$ represent the ferromagnetic triangular Ising model with reinforced interactions, whereas with $J+J_a<0$ we have the antiferromagnetic triangular Ising model which belongs to a frustrated systems class \cite{Moessner2001}.
        \end{itemize}
    \item[II -] $n$ is not a multiple of $6$\label{item2}
        \begin{itemize}
            \item  [ ] we have $J_{ij}=J+J_a$ for connections in the direction of the main axis and $J_{ij}=J-J_a/2$ for the other connections.
        \end{itemize}
\end{itemize}
In this work, we will adopt case \ref{item2} using the value $n=2$. When $J_a=0$, the system represents the triangular Ising model. Otherwise, the interaction energy of a spin with its first neighbors is given by 
\begin{align}\label{math1}
\frac{\mathcal{H}_{i,j}}{\sigma_{i,j}} =& -(1+J_a)(\sigma_{i,j-1}+\sigma_{i,j+ 1})\\\notag &-(1-\frac{J_a}{2})(\sigma_{i-1,j-1}+\sigma_{i-1,j}+\sigma_{i+1,j+1 }+\sigma_{i+1,j}), 
\end{align} 
where the index $i(j)$ is associated with translations in the $\hat{u}$($\hat{v}$) direction. The first term represents the interactions on the main axis, while the second term refers to the interactions outside this axis.
\begin{figure}[!htb] 
\centering \includegraphics[scale=0.75]{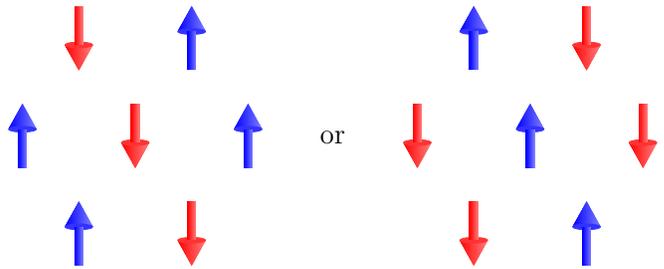} \caption{Ground states for $J_a>2.0$. In the direction of the main axis, the spins have a ferromagnetic order, such configurations are called stripes.\label{fig02}} 
\end{figure}

Para $0<J_a<2$ o sistema se comporta como o modelo de Ising na rede triangular, com estado fundamental composto por configurações ferromagnéticas. When $J_{a}=2$, the interactions outside the main axis are null, making the system analogue to the one-dimensional Ising model with $J=3$. For $J_a>2$ the ground state is formed by stripes ($ST$), which in the direction of the main axis, the spins have a ferromagnetic order, as shown in Fig \ref{fig02}. This happens because the ferromagnetic interaction energy located on the main axis is much greater than the interaction energy of the spins outside this axis.

\section{Computational Details}\label{comp}

We use entropic simulations to obtain the thermodynamic properties of the system to characterize the phase transitions of the model. This tool proved to be effective in the study of critical phenomena due to its ability to obtain thermodynamic quantities at any temperature. Its use became more effective after the publication of the Wang-Landau algorithm \cite{Landau2004, Wang2001}, whereby applying an flatness criterion for the energy histogram, followed at each simulation step, it is possible to obtain a good estimate for the density of states $g(E)$. This methodology allows estimating any thermodynamic quantity $X$ through the canonical mean 
\begin{equation} 
\langle X \rangle =\frac{\sum_{E}X_{E}g(E)e^{-\beta E}}{\sum_{E}g(E)e^{-\beta E}} \label{eq.med}, 
\end{equation}
where $X_{E}$ represents the microcanonical average that was accumulated during the simulation. 

Since the density of states corresponds to a very large number, it is convenient to adhere to the simulation with the logarithm of the density of states, $S(E)=\ln g(E)$, identified as the microcanonical entropy. At the beginning of the simulation, we assume $S(E)=0$ and choose the lowest energy configuration as the starting point. A new configuration is obtained by changing the spin state of a random site, whose acceptance probability is given by 
\begin{equation} 
P(E_\mu \to E_\nu)= min(e^{S(E_\mu)-S(E_\nu)},1). 
\end{equation} 
At each change attempt, we update the energy histogram and the logarithm of the density of states $H(E_\nu) \to H(E_\nu)+1$ and $S(E_\nu) \to S(E_\nu )+ F_i$, respectively. $F_i = \ln f_i$ and $f_i$ the modification factor that initially corresponds to $f_0 \equiv e = 2.71828\dots $\cite{Landau2004}. Then, at each flatness condition met, we update $f_i$ based on the criterion $f_{i+1}=\sqrt f_i$ and then the histogram is reset. Going beyond what is proposed in the original article by Wang-Landau, in this process we accumulate the microcanonical averages from $f_7$ \cite{Caparica2012} where the histogram is expected to be flat. We end the simulation at a $f_{final}$ that ensures the accumulated canonical mean throughout the simulation. In this work we end at $f_{15}$ which is also recognized as the sixteenth level of Wang-Landau. In addition, a two-dimensional approach to the density of states $g(E_1,E_2)$ can be used, allowing estimation of thermodynamic quantities for any values of the energy parameters of $E_1$ and $E_2$ \cite{Ferreira2021}. Its use allows one to obtain a sketch of the phase diagram of the system. However, there is a significant increase in computational time, using only small lattice sizes. The simulations protocol is kept unchanged, the only changes are that $g(E)\longrightarrow g(E_1,E_2)$ and $H(E)\longrightarrow H(E_1,E_2)$. The canonical mean of a quantity $X$ is given by 
\begin{equation}\label{eq.med.bi} 
\langle X \rangle =\frac{\sum_{E_1,E_2}X_{E_1,E_2}g(E_1,E_2)e^{-\beta E}}{\sum_{E_1,E_2}g(E_1,E_2 )e^{-\beta E}}, 
\end{equation} 
where $E=JE_1+J_aE_2$.

\section{Results}\label{results}

To establish a more in-depth description of the behavior of the model, we initially sought to calculate essential thermodynamic quantities such as energy, specific heat, magnetization, susceptibility and other properties through the simulation method linked to the calculation of the density of states $ g(E,M)$. The two-dimensional approach to the density of states allowed us to explore more quickly the influence provoked in the lattice of spins established by the different values assumed by $J_a$. However, despite this gain, we are limited to smaller lattice sizes corresponding to $L=6$, $12$ and $18$. At the points of interest we perform a finite size study using the sizes 30, 36, 42, 48, 54 and 60 with conventional Wang-Landau simulations.

\begin{figure}[!htb]
	\centering
	\includegraphics[scale=0.56,angle=-90]{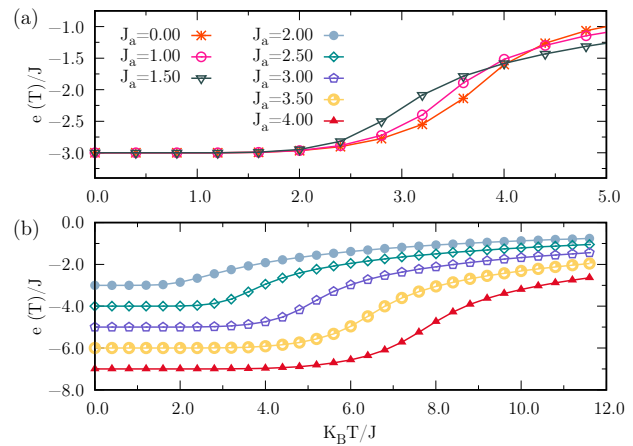}
    \caption{Energy as a function of temperature for several values of $J_a$.}
	%\legend{Fonte: os autores}
	\label{Ene}
\end{figure}

In Fig.\ref{Ene} we obtain the mean energy per spin for $L=12$. We can see that the mean energy at low temperature for $0\leq J_a \leq 2$ is equal to $-3J$, independent of the value $J_a$. This indicates that the ground state is ferromagnetic as analytically expected.

Still in the region of low temperatures when we observe the values of $J_a> 2$, a new symmetry appears in the ground state. We identified as the \emph{stripes} phase (St) which in this case is composed of vertical lines of spins that align in an ordering \textit{up} and \textit{down} alternately, which is also shown according to the analytical exposition. 

\begin{figure}[!htb]
	\centering
	\includegraphics[scale=0.56,angle=-90]{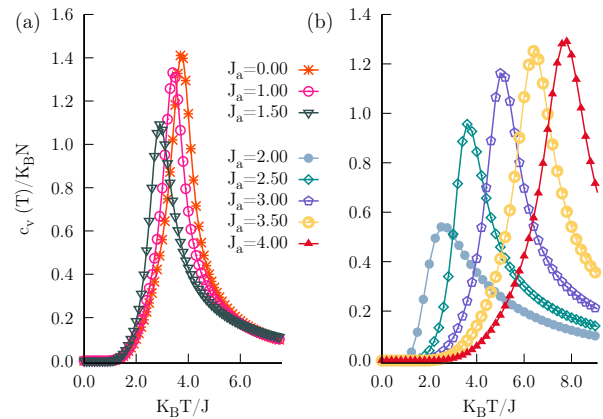}
    \caption{Specific heat as a function of temperature for various values of $J_a$.
}
	%\legend{Fonte: os autores}
	\label{Cv}
\end{figure}

In the Fig.\ref{Cv} we observed the specific heat for values of $J_a$ from 0 to 4. We verify that with $J_a =0.00$ a peak appears corresponding to the critical temperature of the ferro-paramagnetic phase transition, whose analytical value is $k_{B}T_{c}/J \approx 3.65364$ for the pure Ising model in the triangular lattice. Although this value was extracted from a small lattice, it is reasonable as observed in other works \cite{Vatansever2018,Zhi-Huan2009} to provide us with even more confidence regarding the adequacy of pure Ising model.

We observed that the maximum of the specific heat peak undergoes a shift to the left in the range of $0\leq J_a\leq 2$, indicating a progressively lower ferromagnetic transition temperature. In the range of $2<J_a \leq4$, we observed the opposite behavior for the maximum of the specific heat that now undergoes a translation to the right and reaches increasingly higher transition temperatures.

\begin{figure}[!htb]
\centering
 \includegraphics[scale=0.56,angle=-90]{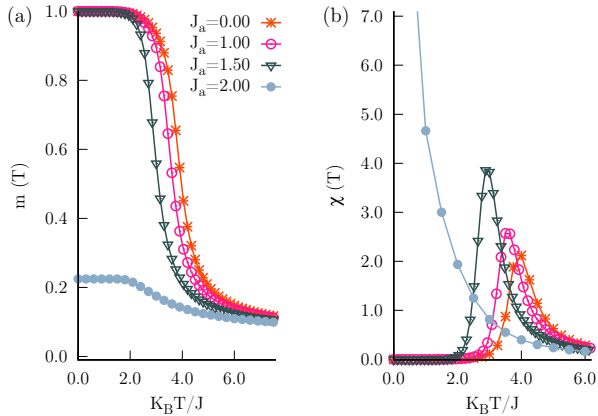}
 \caption{Magnetization as a function of temperature for several values of $J_a$ and magnetic susceptibility as a function of temperature for several values of $J_a$.}
 %\legend{Fonte: os autores}
\label{Mag}
\end{figure}

%The abrupt distinction in behavior reinforces the presence of a new ordered state at low temperatures that appears with increasingly lower energies as shown in Fig.\ref{Ene} for the interval $2< J_a \leq 4$ that need to reach higher energy levels to break the ordering of the \emph{stripes} phase and evolve into the paramagnetic phase. Still from the perspective of the same lattice size,
In Fig.\ref{Mag}a and Fig.\ref{Mag}b we present the magnetization and susceptibility behavior of the system for various values of $J_a$.
The magnetization of the system starts at $1.0$, such as a classic behavior of a ferromagnetic system. The correspondence of these transition points is also observed in the maxima presented by the susceptibility for this same range of $J_a$ values. For $J_a=2.0$, we observe an atypical behavior in both graphs, which was already expected by the analytical calculations of the energy. This value of $J_a$ signals the threshold between two phases. This is the point where the system have one-dimensional dynamics, because of the energetic decoupling of the columns. The one-dimensional behavior is indendified in the specific heat and susceptibility curves \cite{Ferreira2022a}.
\begin{figure}[!htb]
\centering
 \includegraphics[scale=0.56,angle=-90]{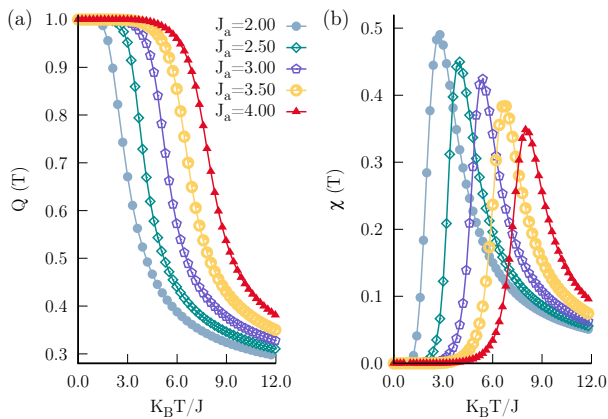}
 \caption{Order parameter and susceptibility as a function of temperature for several values of $J_{a}\geq 2$.}
%\legend{Fonte: os autores}
\label{MagKist}
\end{figure}

To $J_a>2$ the \textit{stripes} phase emerges. To explore this region in more depth, we need to present a new order parameter. This parameter was established as
\begin{equation}
Q=\frac{1}{L^2}\sum_{i=1}^{L}|q_i|,\label{param}
\end{equation}
where $i$ runs over columns and $q_i=\sum_{j=1}^{L}\sigma_{ji}$ is the sum of the spins over the lines in the column $i$. The behavior of this quantity is shown in Fig.\ref{MagKist}a. The susceptibility parameter is obtained by 
\begin{equation}
    \chi_Q=\frac{\langle Q^2\rangle-\langle Q \rangle^2}{k_BT}.
\end{equation}
In Fig.\ref{MagKist}b we show the behavior of this quantity. 

This new order parameter can be used in all range studyed here, because the ground states found here are of the same symmetry of the order parameter. That is, for the ferromagnetic order ground state the sum of the lines is $\pm L$ for any column, the sum of the modulus over all columns is the number of spins of the lattice. Where the behavior is one-dimensional, $q_i$ is the magnetization of lattice, that can be $\pm L$. Again, the sum of the modulus over all columns is the number of spins of the lattice. In the disorder state the all sites have the same probability of spin up and down, so the $q_i\sim 0$.  

%The need to better understand the scalability of the system at points whose regions are well defined together with the guarantee of the validity of the parameter $Q$ raises the confidence level to investigate the \textit{stripes} phase and the ferromagnetic phase for lattice sizes even bigger.

\begin{figure}[!htb]
\centering
\includegraphics[scale=0.56,angle=-90]{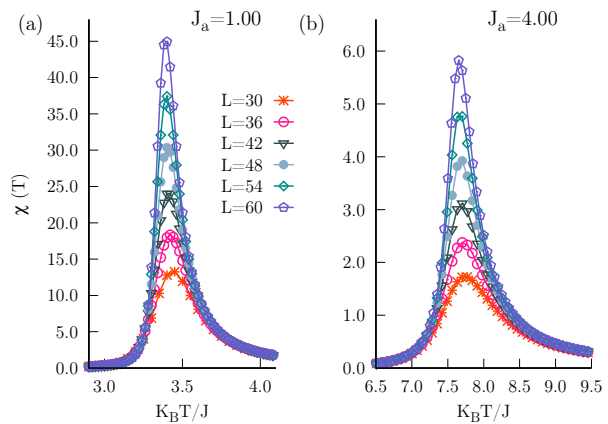}
 \caption{Susceptibility as a function of temperature to $J_a=1.0$. Susceptibility as a function of temperature for $J_a=4.0$}
\label{MaJa2}
\end{figure}

To investigate the scaling law for the order parameter in all regions, we use the $g(E)$ instead of $g(E,M)$ for lattice sizes $30$ to $60$ as can be seen in figures Fig.\ref{MaJa2}a and Fig.\ref{MaJa2}b for. In both graphs, it is possible to see the existence of transition signals for each lattice size and a scaling law that is associated with the respective sizes. 

Since the behavior of the order parameter $Q$ scales with the size of the lattice and the results for the ferromagnetic region are in accordance with what is predicted in the literature, we will adopt the quantity $Q$ as the order parameter of the system in all the regions studied.
\begin{figure}[!htb]
	\centering
    \includegraphics[scale=0.56,angle=-90]{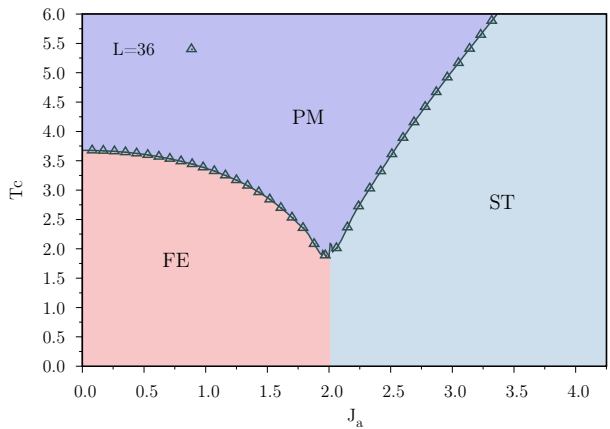}
    \caption{Phase diagram for $J_a\geq0$.}
	%\legend{Fonte: os autores}
	\label{fig03}
\end{figure}

To guide the finite scaling study of this model from the perspective of $J_a>0$, we started our approach at this first moment, seeking to highlight the phase diagram of the model (Fig.\ref{fig03}).
Its construction was established from the extraction of the critical temperature corresponding to the maximum of the specific heat linked to each value attributed to $J_{a}$ that belongs to the interval of interest.
In addition, we also identified that the data shown in Fig.\ref{fig03} has a high degree of similarity with works that, despite having a more complex Hamiltonian, address the same degrees of freedom of spins considered here\cite{Pili2019}.
%In this last reference, for example, it was also observed in the phase diagram of a ferromagnetic system that the coupling with the lattice decreases the transition temperature of the magnetic system before a certain critical coupling value and increases the transition temperature after this value. 

Another aspect that deserves attention regarding this graph is when the model behaves similarly to a one-dimensional system $J_{a}=2.0$. Although the system has a phase transition for $J_a\ne 2$, there is a region close to $J_a\sim 2$ where the system undergoes a change from a two-dimensional system to a one-dimensional system and then an order-to-disorder transition.
\begin{figure}[!htb]
\centering
 \includegraphics[scale=0.56,angle=-90]{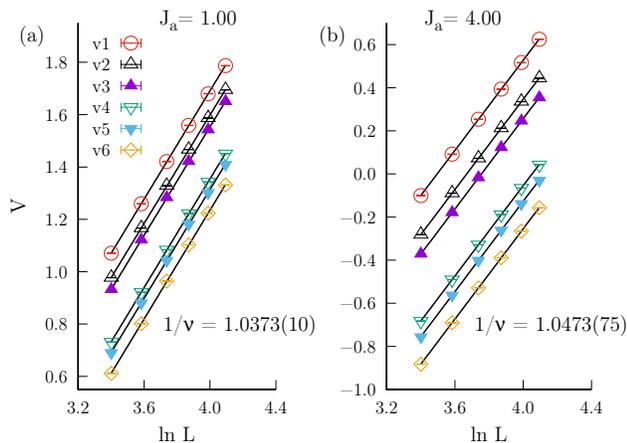}
 \caption{Graphs of $1/\nu$ for $J_a=1.0$ and $J_a=4.0$ respectively.
 }
\label{ni}
\end{figure}
In the two-dimensional system, the study of finite size predicts a power-law scale. In the one-dimensional case, the scale is exponential \cite{Ferreira2023}. In a crossover region these two types of scale will be in competition and a valid theory has not yet been developed. The study of this crossover region is not yet completely understood and therefore we will not delve deeper into the region close to this point.

We use the finite size scaling theory to study the behavior of the system for $J_a=1$ and $J_a=4$, corresponding to the phases ferromagnetic and stripes, respectively. According to this theory we obtain a universal form for the molar Helmholtz free energy corresponding to the equation
\begin{equation}
f(t,H;L)=L^{-d}Y(atL^{1/\nu},bHL^{\Delta/\nu}),
\label{eq.helm}
\end{equation}
$t$ is the reduced temperature, which is equivalent to $(T-T_{c})/T_{c}$, $H$ is the external field, $a$ and $b$ are metric factors, $d$ is the spatial dimension of the system, $\nu$ and $\Delta$ are static critical exponents and $L$ is the linear dimension of the system. From this, it is possible to extract system information equivalent to magnetization, susceptibility and specific heat through the equation (\ref{eq.helm}) \cite{Stanley1971, Caparica2000}. At $H=0$ such features follow the following scaling laws
\begin{equation}
m\approx L^{-\beta/\nu}m'(tL^{1/\nu}) \label{eq.mag},
\end{equation}
\begin{equation}
\chi \approx L^{\gamma/\nu}{\chi}'(tL^{1/\nu}) \label{eq.chi},
\end{equation}
\begin{equation}
c \approx c_{\infty}+L^{\alpha/\nu}{c}'(tL^{1/\nu}).
\end{equation}
\begin{figure}[!htb]
\centering
 \includegraphics[scale=0.56,angle=-90]{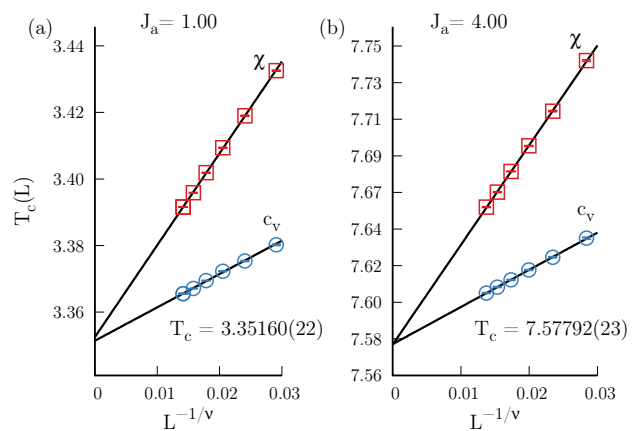}
 \caption{Graphs of $T_{c}$ for $J_a=1.0$ and $J_a=4.0$ respectively. }
%\legend{Fonte: os autores}
\label{Tc}
\end{figure}

In the critical region $(t=0)$, we identify the universal scaling functions $m'$, $\chi'$ and $c'$ as constants and the static critical exponents $\beta$, $\gamma$ and $\alpha$ that make up the base responsible for identifying the universality class that the system fits \cite{Caparica2015b}. These exponents obey the scale and hyperscale relationships recognized as the Fisher, Rushbrooke, Widom, and Josephson \cite{Huang1987} relationships. Despite the usefulness of these relations, it is observed a certain difficulty in eliminating the dependence that each thermodynamic property of interest has on the exponent $\nu$. Thus, to obtain it in isolation, we use the equation 

\begin{equation} V_{j}\approx (1/\nu)\ln{L}+ V'_{j}(tL^{1/\nu}). \label{vjs} \end{equation} 

In this equation we have $j=1,...,6$, where $V'_{j}$ are constants independent of the size of the system that represent thermodynamic quantities extracted from the logarithm of the derivative of magnetization. At the critical temperature, $(T_{c})$ these functions converge to their corresponding value in the infinite lattice so that when we perform a linear adjustment of the graph referring to $V_{j} \times L$ we can calculate $1/\nu $.

Following all the steps of this theory we managed to obtain the first critical exponent of interest as displayed by the graph Fig.\ref{ni}.The results of the critical exponents and critical temperature are presented in Table \ref{tab01}.%Para calcular o erro de $1/\nu$ utilizamos a propagação de incertezas dada por $\Delta\nu =\Delta(1/\nu)/(1/\nu)^2$.

Having found the exponent $\nu$ and the temperature of the specific heat and susceptibility maxima, we can estimate the critical temperature using the equation
\begin{equation}
T_{c}(L) \approx T_{c}+a_{q}L^{-1/\nu},
\label{Tcs}
\end{equation}
\begin{figure}[!htb]
\centering
 \includegraphics[scale=0.56,angle=-90]{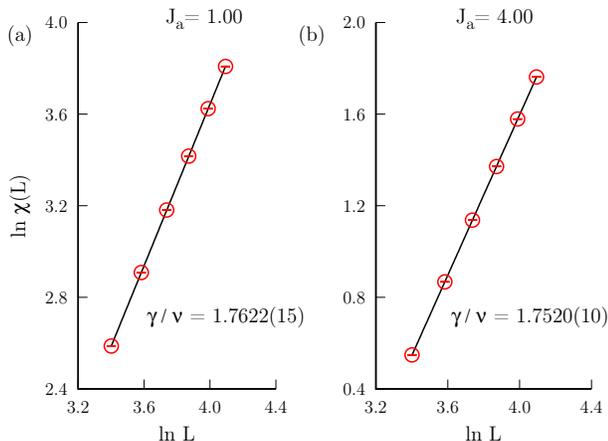}
 \caption{Graph of $\gamma/\nu$ for $J_a=1.0$ and $J_a=4.0$ respectively. }
%\legend{Fonte: os autores}
\label{fg11}
\end{figure}
where $a_{q}$ is identified as a constant. 
\begin{figure}[!htb]
\centering
 \includegraphics[scale=0.56,angle=-90]{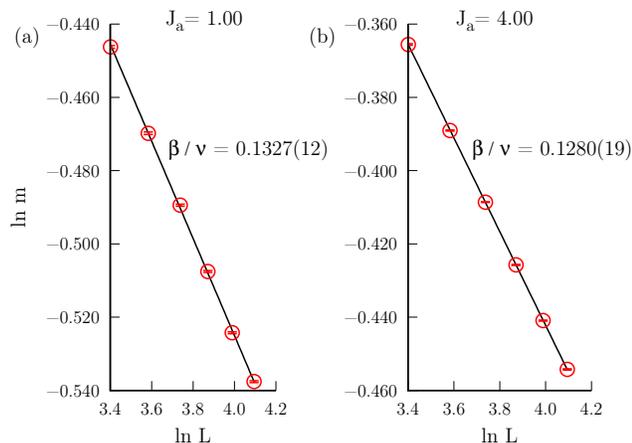}
 \caption{Graph of $\beta/\nu$ for $J_a=1.0$ and $J_a=4.0$ respectively. }
%\legend{Fonte: os autores}
\label{fg12}
\end{figure}
From these data we extract the critical temperature as shown in the graph Fig.\ref{Tc}, in this graph the transition temperature from the ferromagnetic phase ($J_a=1.0$) to the paramagnetic phase is equivalent to $T_{c}=3.35160(22)$, on the other hand, the transition temperature from the stripes phase ($J_a=4.0$) to the paramagnetic phase presents a transition temperature corresponding to $T_{c}=7.57792(23)$.

To estimate the exponent $\gamma/\nu$ we perform a linear adjustment of the log of the maximum susceptibility value by the log of the linear size of the lattice. The results for this procedure are shown in Fig.\ref{fg11}a and Fig.\ref{fg11}b for the cases $J_a=1.0$ and $J_a=4.0$, respectively. We obtain $\gamma=1.6988(31)$ and $\gamma=1.673(13)$ for $J_a=1.0$ and $J_a=4.0$, respectively.

To calculate the critical exponent $\beta$ we perform a linear adjustment of the log of the order parameter at the critical temperature by the log of the linear size of the lattice. The results are shown in Fig.\ref{fg12}a and Fig.\ref{fg12}b for the cases $J_a=1.0$ and $J_a=4.0$, respectively. The values of the critical exponent in each case are $\beta=0.1279(13)$ and $\beta=0.1222(34)$, respectively. The uncertainty of the critical exponents $\gamma$ and $\beta$ were calculated using the uncertainty propagation $\Delta A'=\Delta\nu A+ \nu\Delta A$, where $\Delta A'$ is the uncertainty of the critical exponent and $A$ takes on the values of $\gamma/\nu$ or $\beta/\nu$ . 

This entire set of results just presented for the critical exponents $\gamma/\nu$ and $\beta/\nu$ for the system with $J_a=1.0$ and with $J_a=4.0$ also provides us with strong evidence that the model has the same universality class as the two-dimensional Ising model that has exponents $\gamma/\nu=7/4=1.75$ and $\beta/\nu = 1/8=0.125$ as it is also adopted as a reference parameter in other works \cite{Caparica2014,Vatansever2018}. Furthermore, given the strong evidence exposed so far, it is also possible to infer that the respective transitions from both the ferromagnetic and \textit{Stripes} phases to the paramagnetic phase fall within transitions characterized as second order.

\begin{table}[]
    \centering
    \caption{Estimation of critical exponents and critical temperature.}
    \begin{tabular}{lcccc}
    \hline
         $J_a$ & $1/\nu$     &  $\gamma/\nu $ & $\beta/\nu$  & $T_c$ \\\hline
         $0*$  & ----        &  1.75          &  0.125       &  3.65364      \\
         $1$   & 1.0373(10)  &  1.7622(15)    &  0.1327(12)  &  3.35160(22)\\
         $4$   & 1.0473(75)   &  1.7520(10)    &  0.1280(19)   & 7.57792(23) \\ \hline\hline
    \end{tabular}

    \label{tab01}
\end{table}

\section{Conclusion\label{conclusion}}

The finite-size scaling study showed that this model is in the same universality class as the Ising model, with critical exponents presented in TABLE.\ref{tab01}, showing that the influence of the lattice vibration on the exchange interaction promotes the change of fundamental state, but preserves the class of universality. We also observed an increase in the transition temperature for the stripes region compared to the ferromagnetic region.

Finally, we suggest that a detailed study be carried out to investigate the region in which the system stops being two-dimensional and becomes one-dimensional, since the crossover region between two scaling laws is not well established in literature. Such a study can glimpse important characteristics for the study of models where there is a need for corrections to the scaling law in the form of critical temperature power.
\bibliography{referencias.bib}

\end{document}